\newcommand{\mnras}[1]{MNRAS}
\newcommand{\apj}[1]{ApJ}
\newcommand{\apjs}[1]{ApJS}
\newcommand{\apjl}[1]{ApJL}
\newcommand{\nat}[1]{Nature}
\newcommand{\aap}[1]{A\&A}
\newcommand{\araa}[1]{ARA\&A}
\newcommand{\aaps}[1]{A\&ASS}
\newcommand{\aj}[1]{AJ}
\newcommand{\apss}[1]{Ap\&SS}

\documentstyle[epsf]{mn}

\title[Microlensing in GRB Afterglows]{On the Probability of Microlensing in GRB Afterglows}

\author[L.V.E.\,Koopmans \& J. Wambsganss]{L.V.E.\,Koopmans$^1$ \& J.\,Wambsganss$^{2,3,4}$\\
 $^1$California Institute of Technology, mailcode 130-33, Pasadena CA 91125, US\\
 $^2$Universit\"at Potsdam, Institut f\"ur  Physik, Am Neuen Palais 10, 14469 Potsdam, Germany\\
 $^3$Max-Planck-Institut f\"ur Gravitationsphysik, ``Albert-Einstein-Institut", Am M\"uhlenberg 1,
     14476 Golm, Germany\\
 $^4$University of Melbourne, School of Physics, Parkville, Vic 3052, Australia}

\date{Accepted ... Received ...}
\pubyear{2001}

\begin{document}

\label{firstpage}

\maketitle

\begin{abstract}

The declining lightcurve of the optical afterglow of gamma-ray burst
GRB000301C showed rapid variability with one particularly bright
feature at about $t$$-t_0$=3.8 days.  This event was interpreted as
gravitational microlensing by Garnavich, Loeb \& Stanek (2000) and
subsequently used to derive constraints on the structure of the GRB
optical afterglow. In this paper, we use these structural parameters
to calculate the probability of such a microlensing event in a
realistic scenario, where all compact objects in the universe are
associated with observable galaxies.  For GRB000301C at a redshift of
$z$=2.04, the {\sl a posteriori} probability for a microlensing event
with an amplitude of $\Delta m$$\ge$0.95\,mag (as observed) is 0.7\%
(2.7\%) for the most plausible scenario of a flat $\Lambda$-dominated
FRW universe with $\Omega_{\rm m}$=0.3 and a fraction $f_*$=0.2 (1.0)
of dark-matter in the form of compact objects. If we lower the
magnification threshold to $\Delta m$$\ge$$0.10$\,mag, the
probabilities for microlensing events of GRB afterglows increase to
17\% (57\%). We emphasise that this low probability for a microlensing
signature of almost a magnitude does {\sl not} exclude that the
observed event in the afterglow lightcurve of GRB000301C was caused by
microlensing, especially in light of the fact that a galaxy was found
within 2~arcsec from the GRB. In that case, however, a more robust
upper limit on the {\sl a posteriori} probability of $\approx$5\% is
found. It does show, however, that it will not be easy to create a
large sample of strong GRB afterglow microlensing events for
statistical studies of their physical conditions on micro-arcsec
scales.

\end{abstract}

\begin{keywords}
 cosmology: gravitational lensing -- dark matter -- gamma rays: bursts
\end{keywords}

\section{Introduction}

Gravitational microlensing offers a way to study the structure of high
redshift sources on micro-arcsecond scales, besides being able to
constrain the mass fraction and mass function of compact objects in
the universe. In strong gravitational lens systems (i.e. systems with
multiple images of a single background source), the microlensing
optical depth is of order unity (e.g. Chang \& Refsdal 1979; Gott
1981; Young 1981). Precisely because of this high optical depth, these
systems are perfect for resolving micro-arcsec structure in the lensed
cosmologically-distant source if it crosses a caustic created by the
stellar-mass compact objects in the lens mass distribution (e.g. Chang
\& Refsdal 1984; Grieger, Kayser \& Refsdal 1988; Wambsganss,
Paczy\'nski \& Schneider 1990; Wo\'zniak et al. 2000). Moreover,
because of the presence of multiple images, one can in principle
separate intrinsic source fluctuations from microlensing variability.

For example, ongoing microlensing of the lensed (optical) images in
the system Q2237+0305 (Huchra et al. 1985) has been observed ever
since its discovery (e.g. Irwin et al. 1989; Corrigan et al. 1991;
{\O}stensen et al. 1996; Lewis et al. 1998; Wo{\'z}niak et al. 2000).
To a smaller degree and on longer time scales, microlensing in
Q0957+561 has been detected as well (Pelt et al. 1998; see also
Refsdal et al. 2000). The time scale of microlensing in both cases is
defined by the relative transverse velocity between the source, lens
and observer, which is given by the bulk velocity of the lensing
galaxy plus the random motions of compact objects in the line-of-sight
to the stationary quasar images, and is typically of order several
hundred km\,s$^{-1}$. This results in microlensing time scales of the
order of months to years for solar-mass objects (e.g. Wambsganss
2000).

%
%
% FIGURE 1:
%
%
\begin{figure*}
\begin{center}
\leavevmode
\hbox{%
\epsfxsize=\hsize
\epsffile{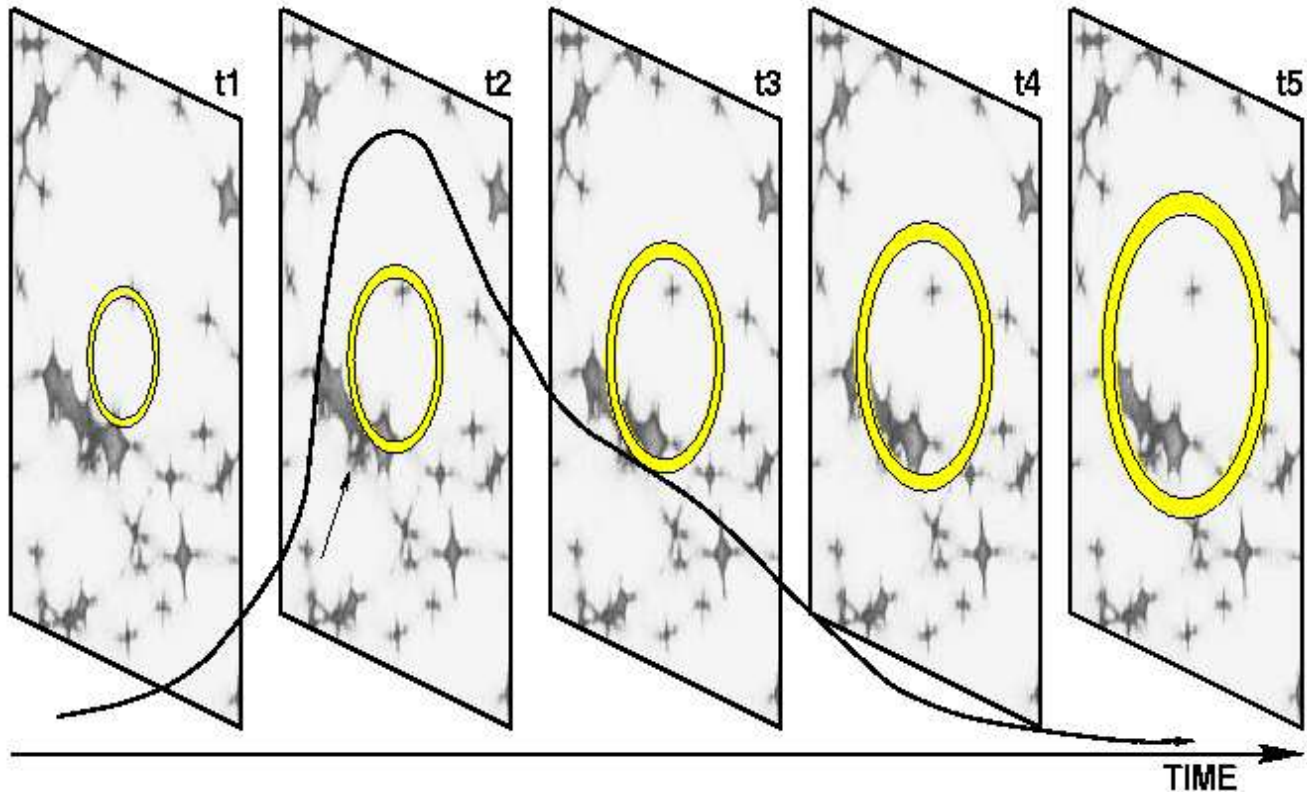}}
\end{center}
\caption{A cartoon of how we determine the microlensing magnification
as a function of time for the expanding shell source superimposed on the
magnification patterns caused by massive compact objects. The five panels
indicate five different epochs (in practise, we consider 100 epochs).
The superimposed curved line indicates the microlensing magnification 
as a function of time, corresponding to this particular example.
The arrow points to a particular high magnification structure that is
passed by the ring source at epoch $t_2$.}
\end{figure*}

Besides these optical sources, recently the first case of
radio-microlensing was reported in the lens system B1600+434 (Koopmans
et al. 2000a, 2000b; Koopmans \& de Bruyn 2000), suggesting extremely
compact relativistic substructure in the lensed radio source. If this
microarcsec-scale substructure is part of a relativistic jet, the
time-scale of microlensing variability by stellar-mass objects reduces
to several weeks (Koopmans \& de Bruyn 2000), allowing one to probe
compact objects up to $\sim$$10^{5}$\,M$_\odot$ on time scales of
several years, as well as the substructure of the relativistic jet.
Indications of optical microlensing in B1600+434 with similar time
scales have also been found (Burud et al. 2000). Clearly, microlensing
is a promising field of future research regarding the study of
high-$z$ sources at micro-arcsecond scales.

In addition to testing the structure of AGNs, Loeb \& Perna (1998)
more recently proposed the use of microlensing to probe into the
internal structure of GRB afterglows on micro-arcsec scales.
Garnavich, Loeb \& Stanek (2000) indeed interpreted an anomalous event
in the lightcurves of the optical afterglow of GRB000301C (Masetti et
al. 2000; Sagar et al. 2000; Berger et al. 2000; Jensen et al. 2000;
Smette et al. 2000) as being caused by microlensing of the GRB
afterglow. They subsequently derive constraints on its structure,
which appear in good agreement with theoretical blastwave models. The
inferred mass of the lensing object is $\sim$0.5\,M$_\odot$, if its
redshift is optimal for microlensing (i.e. about half way; Garnavich
et al. 2000).

However, one needs to be cautious here, because no multiple images are
present -- as in the case of strong gravitational lenses -- to confirm
that this is indeed a non-intrinsic event. The fact that the event
occurs within only a few days after the burst and has an amplitude of
$\approx$1 mag, suggests that the burst must have occurred close to
the Einstein radius of an intervening massive compact object
(Garnavich et al. 2000). To have a significant probability of
observing such a GRB microlensing event, the universe requires a
surface density in compact objects close to the critical surface
density (e.g. Press \& Gunn 1973; Blaes \& Webster 1992).  The
situation is partly similar to the case of variability in single
quasars (i.e. not multiply-imaged), which also cannot easily be proven
to be due to microlensing (e.g. Hawkins \& Taylor 1997; Hawkins 1998),
although the freedom to model the GRB afterglow lightcurve (mostly
dominated by self-similar expansion resulting in a power-law behavior)
is significantly less than that for quasars.

In this paper, we investigate the probability of microlensing in GRB
afterglows for a flat $\Lambda$--dominated cosmological model and with
the distribution of massive compact objects connected to visible
galaxies, particularly focusing on the strong event seen in the
GRB000301C afterglow. Our conclusions, however, are independent of
this particular GRB. In Section 2, we describe numerical microlensing
simulations of GRB afterglow light curves, extended to high
microlensing optical depths. In Section 3, the {\sl a posteriori}
probability of the observed event in GRB000301C is calculated. Section
4 summarizes our result and states our conclusions.

\section{Microlensed GRB Light Curves}

To calculate the microlensing probability of GRBs, we simulate
microlensing magnification patterns for a range of shears and
dimensionless surface densities in compact objects (i.e. $\gamma$ and
$\kappa$; see Schneider, Ehlers \& Falco 1992, Chap.5 for the
definitions).  A similar analysis for the microlensing variability in
Q2237+0305 was done by Rauch \& Blandford (1991) and Jaroszynski et
al. (1992), placing constraints on its accretion disk models.

As a first-order model, we here assume that (i) galaxies in the
universe can be described as a singular isothermal sphere mass
distribution (SIS; e.g. Binney \& Tremaine 1987) for which
$\kappa$=$\gamma$ (e.g. Kormann et al. 1994), (ii) the fraction of
mass in compact objects $f_*$=1, although in Sect.3.3 we will consider
the case of $f_*$$<$1, and (iii) the mass spectrum of compact objects
is narrow ($\Delta m/m$$<$1) such that we can assume that all objects
have nearly the same mass (for example the average or median value).
Magnification patterns for $\kappa$=$\gamma$=0.005, 0.01, 0.025, 0.05,
0.1 and 0.25 are calculated on a grid of 1024$\times$1024 pixels using
the ray-shooting algorithm of Wambsganss (1999). Each pixel has a size
of 0.1~Einstein radius.

For the structure of the GRB, we take the results from Garnavich et
al. (2000), who model the GRB as an expanding ring with a width $W$
times the ring radius $R(t)$. No jet structure is assumed in the
model. The ring radius evolves as function of time as $R(t)=R_0\times
t^{5/8}$ (e.g. Waxman 1997), where $t$ is time in days and $R_0$ is
the ring radius on day one. Their best fit to the combined optical
data sets suggests $W$=0.16$\pm$0.02 and $R_0$=0.49$\pm$0.02 Einstein
radius. A constant surface brightness inside the ring is assumed and
no emission comes from either inside or outside of the ring structure.

Although the redshift and mass of the possible lensing object is
unknown, this is of no consequence to our calculations, because we can
express the properties of lensing objects in dimensionless units of
critical surface density, shear and Einstein radius. For intermediate
redshifts the mass of the possible lensing object of the GRB000301C
afterglow would be $\sim$0.5\,M$_\odot$ (Garnavich et al. 2000), which
agrees well with Galactic (e.g. Alcock et al. 2000) as well as
extragalactic constraints on halo mass objects (Refsdal et al. 2000;
Wambsganss et al. 2000).

To obtain simulated GRB microlensing light curves, we convolve the
magnification patterns with the time-variable GRB source structure for
each epoch $t$.  We then store the `convolved' magnifications from
$10^4$ random grid points.  We repeat this procedure for 100 epochs
from day 0.5 to day 35 (approximately the period during which the
afterglow lightcurve of GRB000301C was sampled; Garnavich et
al. 2000), each step evolving the GRB afterglow structure and taking
the magnifications from the same $10^4$ grid points (the procedure is
illustrated in Figure 1). The sampling epochs for our simulated
microlensing lightcurves are logarithmically spaced, to avoid
undersampling during the initial GRB afterglow phase, where one might
expect the highest magnifications.  The light curves are normalized to
show only the microlensing magnification of the GRB afterglow (as in
Mao \& Loeb 2000).  In Figure 2, we show the probability density
distributions of the peak magnification of the normalized GRB
microlensing light curves, from which it is immediately clear that
strong $\approx$1\,mag events are relatively rare. We will discuss
this in more detail in the next section.

\begin{table}
\centering
\begin{tabular}{rcc}
\hline
 $\kappa$=$\gamma$ & P($\Delta m $$\ge$0.95$^{\rm m}$) &
	P($\Delta m $$\ge$0.10$^{\rm m}$)\\
\hline
 0.005& 0.0043$\pm$0.0007 & 0.181$\pm$0.004\\ 
 0.010& 0.0063$\pm$0.0008 & 0.224$\pm$0.005\\ 
 0.025& 0.0150$\pm$0.001 & 0.441$\pm$0.006\\ 
 0.050& 0.0034$\pm$0.002 & 0.617$\pm$0.008\\ 
 0.100& 0.0530$\pm$0.002 & 0.729$\pm$0.009\\ 
 0.250& 0.0530$\pm$0.002 & 0.781$\pm$0.009\\
\hline
\end{tabular}
\caption{The {\sl a posteriori} probability that GRB000301C shows a microlensing 
event with magnitude greater than approx. 0.10 and 0.95\,mag,
respectively. The errors indicate the Poisson error, due to the finite
number (i.e. $10^4$) of simulated light curves. The equal probability
for $\kappa$=0.1 and 0.25 in the second column is coincidental, but
illustrates the rapid break from $P$$\propto$$\kappa$ (see text) in
the probability of strong events for high optical depth regimes.}
\end{table}

\section{Probabilities}

In this section we consider the probability of observing the event
seen in GRB000301C, within the microlensing hypothesis. First, we
calculate the probability of the event as function of the
dimensionless surface density in compact objects. Second, we determine
the probability of actually observing GRB000301C through a
dimensionless surface density, $\kappa$, as function of a cosmological
model and distribution of compact objects. Third, we combine these
probabilities to arrive at an {\sl a posteriori} probability for the
feature in the afterglow light curve of GRB000301C being a
microlensing event, as suggested by Garnavich et al. (2000).

\subsection{Event-Amplitude Probability}

The microlensing event seen in GRB000301C has a maximum amplitude of
0.95 mag or a magnification $\mu_{\rm o}$=2.4 (see Figure 2 in
Garnavich et al. 2000). To calculate the probability of at least such
a strong event, we need to measure the fraction of simulated light
curves (Sect.2) that show an event stronger than $\mu_{\rm o}$ (see
Figure 2). These numbers are listed in Table~1. We find that even in
the higher optical depth regimes the probability does not
significantly exceed $\approx$5\%, but levels off in the high optical
depth regime, because of averaging of the GRB afterglow over multiple
caustics. At low surface densities, one expects the probability of
microlensing ($\tau_{\mu{\rm l}}$) to asymptotically behave as
$\tau_{\mu{\rm l}}$$\propto$$\kappa$.

Let us now make a simple analytical fit through these probabilities,
using the following function
\begin{equation}	
	P_A(\kappa)\approx\frac{P_{\rm c}\, \kappa}
	{\sqrt{1+(\kappa/\kappa_{\rm c})^2}}.
\end{equation}
This function is by no means the correct functional form, that might
be expected from a detailed theoretical analysis, but it has the
correct asymptotic behaviour for $\kappa$$\rightarrow$0. We find that
$P_{\rm c}$$\approx$0.82 and $\kappa_{\rm c}$$\approx$0.072 fit the
results in Table~1 best. For the purposes here, this empirical
function provides a good enough representation of the probability of
the observed event in the region $0 < \kappa \le 0.25 $. The value
$\kappa$=0.25, is the highest through which any source can be seen
without being multiply-imaged in case of a SIS mass distribution
(e.g. Kormann et al. 1994).  In Table~1, we also list the results for
events with amplitudes greater than 0.1\,mag (i.e. $\mu_{\rm
o}$=1.1). In that case, we find $P_{\rm c}$$\approx$22.7 and
$\kappa_{\rm c}$$\approx$0.034. We defer a discussion of these events
to Sect.3.3.

For the high-magnification events ($\ge$0.95\,mag), Table~1 shows that
$P$$\la$$\kappa$. Because the probability of multiple imaging by
compact objects is equal to $\kappa$ (in the case of $\kappa$$\ll$1),
this result implies that the GRB needs to lie very close to the
Einstein radius of the compact object as found for GRB000301C
(Garnavich et al. 2000). Conversely, one could conclude: The fact that
the microlensing event in GRB000301C requires the GRB to lie close to
the Einstein radius, implies a probability of the event close to the
average dimensionless surface density $<$$\kappa$$>$ of the universe
in compact objects.

%
%
% FIGURE 2:
%
%
\begin{figure*}
\begin{center}
\leavevmode
\hbox{%
\epsfxsize=\hsize
\epsffile{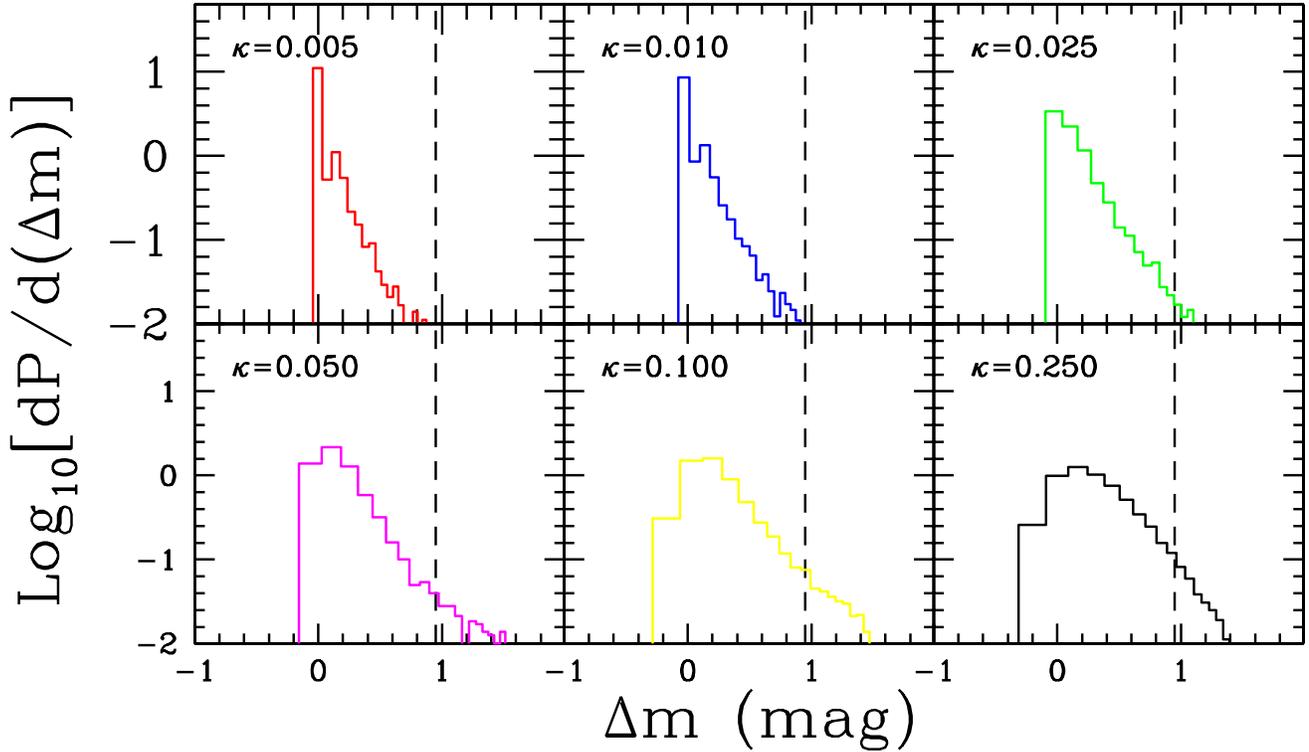}}
\end{center}
\caption{The probability distribution function of microlensing magnifications
as function of $\kappa$=$\gamma$. From upper left to lower right,
$\kappa$=$\gamma$=0.005, 0.01, 0.025, 0.05, 0.1 and 0.25,
respectively.  The dashed line indicates a peak magnification of
0.95~mag or $\mu_{\rm o}$=2.4.}
\end{figure*}

\subsection{Surface Density Probability}

To obtain the final probability of this event being observed, we have
to multiply eqn.(1) with the probability that the GRB is seen through
a region of sky with a dimensionless surface density between $\kappa$
and $\kappa$+$d\kappa$.

For a constant comoving density of SIS lens galaxies, which
distribution follows the Schechter luminosity (i.e. mass) function, we
find from Turner, Ostriker \& Gott (1984), Turner (1990) and Fukugita
\& Turner (1991) that the optical depth for multiple imaging is
\begin{equation}
	\tau_{\rm GL}(z_{\rm s}) = \frac{16 \pi^3}{30} n^*_0
	\left(\frac{c}{{\rm H}_0}\right)^3
	\left(\frac{\sigma^*_{||}}{c}\right)^4
	\Gamma\left[\alpha+\frac{4}{\gamma}+1\right]\,G(z_{\rm s}),
\end{equation}
where $n^*_0$ and $\sigma^*_{||}$ are the local density and central
velocity dispersion of $L_*$ galaxies, respectively. The parameters in
the $\Gamma$--function describe the Schechter luminosity function (see
Fukugita \& Turner (1991) for a detailed description of this equation).
Furthermore
\begin{equation}
	G(z_{\rm s})=\left[\int^{1+z_{\rm s}}_1
	\frac{dw}{\sqrt{\Omega_{\rm m} w^3 - \Omega_{\rm
	m}+1}}\right]^3,
\end{equation}
for a flat FRW universe with $\Omega_{\rm m}+\Omega_{\Lambda}$=1 and a
GRB redshift $z_{\rm s}$. For definiteness, we assume here that
$\Omega_{\rm m}$=0.3 and $\Omega_\Lambda$=0.7, which seem to agree
best with most recent CMB, SNe Ia and cluster-abundance observations.

The optical depth $\tau_{\rm GL}(z_{\rm s})$ indicates the fraction of
sky covered by regions in which sources are multiply imaged
(i.e. regions inside the Einstein ring). It is easy to show that the
brightest image, in case of multiple imaging of the source, always
lies in the region $\kappa$=0.25--0.50 (between 1--2 Einstein radii
from the lens centre) for a SIS mass distribution ($\kappa$=$1/2x$,
with $x$ being the radius in Einstein radii). This region projects
one-to-one back onto a disk of one Einstein radius in the source plane
(see Schneider et al. 1992, Chapter 8). Hence the probability of
observing a GRB through a patch of sky with dimensionless surface
density $>$$\kappa$ (but smaller than $\kappa$=0.5) is given by
\begin{equation}
	P(>\kappa)=\tau_{\rm GL}(z_{\rm s})\times\left(\frac{1}{2\kappa} -1 \right)^2,
\end{equation}
in which case the region $\kappa$=0.25--0.50 exactly has the
probability $\tau_{\rm GL}(z_{\rm s})$, as required. We do not take
the magnification bias into account here, because the majority of GRB
will be seen through regions with $\kappa$$\ll$1. In Sect.4, we will
come back to the magnification bias in more detail. Finally, we need
to normalize $P(>\kappa)$ to unity, which requires us to put a lower
limit on the allowed $\kappa$. We find $\kappa_{\rm
l}$=$\frac{1}{2}(1+ 1/\sqrt{\tau_{\rm GL}})^{-1}$ for which
$P(>\kappa_{\rm l})$=1.  For surface densities smaller than
$\kappa_{\rm l}$, galaxies will start to significantly overlap. In
that case eqn.(4) will clearly break down and $\kappa_{\rm l}$ is
therefore a natural limit on the probability distribution of $\kappa$.

\subsection{Probability of the GRB000301C Event}

The overall probability that a GRB with the properties of
GRB000301C shows a microlensing event with an amplitude
of $\approx$0.95 mag (i.e. $\mu_{\rm o}$=2.4) then becomes
\begin{equation}
	P_{\rm e}=\int_{0.5}^{\kappa_{\rm l}} \frac{d P(>\kappa)}{d
	\kappa} \times P_{\rm A}(\kappa\, |\, \mu_{\rm o})\, d\kappa.
\end{equation}
If we now use the values $n_0^*$=0.61$\times10^{-2}h^3$Mpc$^{-3}$,
$\alpha$=$-$1, $\gamma$=4 and $\sigma_{||}^*$=225 km/s for the
Schechter luminosity function describing the population of elliptical
lens galaxies, which presumably dominates the lensing cross--section
and mass in the universe (see also Kochanek 1996; Falco et al. 1998),
we find
\begin{equation}
	\tau_{\rm GL}(z_{\rm s}) \approx 9\cdot10^{-4}\times G(z_{\rm
	s}).
\end{equation}
Evaluating eqn.(6) for the redshift of the burst $z_{\rm GRB}$=2.04
(Jensen et al. 2000; see also Smette et al. 2000), we find the
probability that GRB00301C could have been multiply imaged to be
\begin{equation}
	\tau_{\rm GL/GRB}\approx2.2\times 10^{-3},
\end{equation}
for $\Omega_{\rm m}$=0.3 in a flat FRW universe. Hence the probability
of seeing the observed microlensing event becomes
\begin{equation}
	P_{\rm e} \approx \tau_{\rm GL/GRB} \times
	\int_{0.5}^{\kappa_{\rm l}} \frac{\left( 2\,\kappa -1 \right)
	\,P_{\rm c}} {2\,\kappa^2\,{\sqrt{1 + \left(\kappa/\kappa_{\rm
	c} \right)^2}}}.
\end{equation}
For GRB000301C at $z_{\rm z}$=2.04, we furthermore find $\kappa_{\rm
l}$$\approx$0.022 (Sect.3.2). Evaluating eqn.(8) for $P_{\rm
c}$$\approx$0.82 and $\kappa_{\rm c}$$\approx$0.072 (Sect.3.1), we
finally find that the {\sl a posteriori} probability of the event seen 
in GRB000301C is $P_{\rm e}(f_*=1) \approx 0.027$.  

However, this calculation assumes that all mass is in the form of compact
objects.  Recent results from the MACHO collaboration indicate that
$f_*$=0.08--0.50 with 95\% confidence (Alcock et al. 2000).  The most
likely value is $f_*$$\approx$0.2. In that case, we have to modify
eqn.(1) such that $\kappa$$\rightarrow$$f_*\,\kappa$. In other words,
the probability of an event goes down by a factor $\sim$$f_*$.  This
scaling of the numerical simulation is allowed, because for
$\kappa$=$\gamma$$\ll$1, the influence of the shear on lensing
properties is small (see also Mao \& Loeb 2000). Similarly, changing a
fraction $(1-f_*)$ of the surface density in compact objects to a
smooth mass distribution is nearly similar to completely removing this
fraction for $\kappa$$\ll$1. Thus for a more realistic scenario, the
probability reduces to $P_{\rm e}(f_*=0.2)$$\approx$0.007.

For microlensing events with an amplitude $\Delta m
\ge 0.1$mag (much weaker events than the one seen in the GRB0003001C
afterglow), we find $P_{\rm e}$$\approx$0.56 and 0.17, respectively,
for $f_*$=1.0 and 0.2.  One might therefore expect many high-$z$
(i.e. $z$$\ga$2) GRB afterglows to show evidence for microlensing at a
low level, assuming their physical properties are similar to those 
inferred from GRB000301C by Garnavich et al. (2000).

\section{Discussion}

If the event seen in the optical lightcurve of the GRB000301C
afterglow is caused by microlensing, the opportunities to study the
structure and evolution of GRB afterglows on micro-arcsec scale are
potentially very exciting (Loeb \& Perna 1998; Garnavich et
al. 2000; Mao \& Loeb 2000).

However, in this paper we have shown that the {\sl a posteriori}
probability of this particular event is actually small,
i.e. 0.7--2.7\% for a fraction $f_*$=0.2--1.0, respectively, of
dark-matter in the form of compact objects. The main assumptions in
this calculation are: (i) a constant comoving density population of
galaxies which follow a Schechter luminosity (i.e. mass) function,
(ii) a flat $\Lambda$--dominated FRW universe with $\Omega_{\rm
m}$=0.3, (iii) all matter in the universe traces galaxies, which can
be described as singular isothermal spheres, (iv) the mass spectrum of
compact objects is `narrow', in which case they can be parameterised
by a single value for their mass, (v) a typical GRB optical afterglow
has similar properties as GRB000301C and has a redshift of
$z$$\approx$2, and (vi) there is no significant magnification bias.
The magnification bias, however, is unlikely to be a problem. Even in
case this bias increases the number of observed GRBs by a factor of 10
for high microlensing optical depths ($\kappa$$\ga$0.25), Table~1 and
eqn.(7) show that the overall probabilities are increased by only
10$\times P(\Delta m \ge 0.95^{\rm m}|\kappa\ga0.25)
\times\tau_{\rm GL/GRB}$$\sim$0.1\%. A detailed calculation should
consistently take into account both the magnification distribution of
galaxies {\sl and} that of compact objects (e.g. Pei 1993a, 1993b). In
light of the very uncertain redshift distribution and luminosity
function of GRB afterglows this is, however, not yet warranted.

Moreover, even in the case that the {\sl whole} sky has a surface mass
density of $\kappa_*$$\la$0.25 (normalised by the ``critical'' surface
mass density for lensing, e.g. Schneider et al. 1992) our simulations
(see Table~1) indicate that the probability of such an event does not
significantly exceed $\approx$5\%.  Clearly, the latter situation is
unrealistic, but it does place a very robust upper limit on the
probability of this event (see also Press \& Gunn 1973 and Blaes \&
Webster 1992).

The plausibility to actually observe such a fraction of GRB afterglow
lightcurves with a microlensing event of about one magnitude also depends on
the typical mass of compact objects.  However, even if the mass goes
up or down by a factor of ten from the assumed 0.5\,M$_\odot$, to
first approximation that would just mean a ``broadening'' in time of
the microlensing event by about a factor of three (since the
length/time scale is proportional to the square root of the mass).
However, if most compact objects have masses {\it much} smaller than
that inferred for GRB000301C, the fraction of GRB afterglows with
microlensing signatures will obviously decrease. In that case the
effect of the ``shrunk'' caustics will average out over the much
larger angular size of the emission region of the GRB afterglow. This
case, however, seems unlikely in light of the stringent lower limit
($\sim$10$^{-2}$\,M$_\odot$) placed on low-mass compact objects in
0957+561 (Schmidt \& Wambsganss 1998; Refsdal et al. 2000; Wambsganss
et al. 2000). The fraction of microlensed GRB afterglows would also
decrease if the typical redshift of GRBs is significantly lower
than~$z=2$. A somewhat lower typical redshift is supported by the
average redshift $<z>_{\rm GRB}\approx1.3$, with an rms spread of
$\approx$1, that we find from Bloom, Kulkarni, \& Djorgovski (2000).
In case of a broad mass spectrum of the compact objects or a large
spread in the physical properties of the GRB optical afterglows, one
has to convolve the microlensing probabilities with these properties.
This, however, is not expected to change the results significantly.

We emphasise that our results {\sl do not exclude} that the event seen
in the lightcurve of the optical afterglow of GRB000301C is in fact
due to gravitational microlensing, especially in light of the fact that
a galaxy was found at 2~arcsec from the GRB (Garnavich et al. 2000).
Each event should be treated on its own merit (and it is a posteriori
anyway). In any case, in the sample of $\sim$20 known GRBs with
observed optical afterglows (e.g. Bloom et al. 2000), the probability
of one such event would be around 14\%.  But our results do predict
that it is unlikely to find many strongly microlensed GRB afterglow soon,
and it emphasises the difficulties one will encounter in creating a
large sample of these strong (i.e. $\Delta m \ga1$\,mag) events,
with the aim to statistically study the physical properties of GRB
afterglows on micro-arcsec scales.  For example, to get 10 additional
microlensing events of comparable strength, one would require
$\approx$1500 GRB afterglow optical lightcurves. Even for the {\sl
Swift} satellite (Parsons et al. 1999; see also Mao \& Loeb 2000) this
provides a challenging task.

On the bright sight, if the event in GRB000301C was due to
microlensing, we can expect a significant fraction of high-$z$ GRB
afterglow light curves to show microlensing events stronger than
0.1\,mag (see also Mao \& Loeb 2000), although these will be hard to
distinguish from intrinsic variations and will also not really be able
probe the GRB afterglow structure on micro-arcsec scales. We conclude,
that the best cases to unambiguously study micro-arcsec structure in
GRB afterglows will probably be those GRBs that are multiply imaged
(about 1--2\% expected, see, e.g.  Holz et al. 1999; Komberg et
al. 1999; Mao 1992, 1993; Marani et al. 1999; McBreen et al. 1993;
Narayan \& Wallington 1992; Nowak \& Grossman 1993; Paczy\'nski 1987;
Wambsganss 1993; Williams \& Wijers 1997).  Those (soon to be
discovered!) will almost certainly (see Table~1) show microlensing
events $\ga$0.1\,mag, which can easily be separated from intrinsic
variations by comparison with the other lensed GRB images, after
correcting for the respective time delays. A smaller fraction
($\approx$5\%) will also show events $\ga$1\,mag, although this might
increase due to the magnification bias. So the prospects of
doing science with (micro-)lensed gamma ray bursts and afterglows
remain promising.

\section*{Acknowledgements}

The authors would like to thank Roger Blandford, Avi Loeb and Shude Mao 
for useful comments. 

{}

\end{document}